\documentclass[12pt,letterpaper]{article}          
\usepackage{osajnl}
\usepackage[draft]{hyperref} 

\begin{document}

\title{Gaussian and non-Gaussian speckle fluctuations in
the diffusing-wave spectroscopy signal of a coarsening foam}

\author{A.S. Gittings$^1$ and D.J. Durian$^{1,2}$}

\address{$^1$University of California, Department of Physics \& Astronomy,
Los Angeles, CA 90095}
\address{$^2$University of Pennsylvania, Department of Physics \& Astronomy,
Philadelphia, PA 19104}

\email{djdurian@physics.upenn.edu}

\begin{abstract}
All prior applications of Diffusing-Wave Spectroscopy (DWS) to
aqueous foams rely upon the assumption that the electric field of
the detected light is a Gaussian random variable and that, hence,
the Siegert relation applies.  Here we test this crucial assumption
by simultaneous measurement of both second and third-order temporal
intensity correlations.  We find that the electric field is Gaussian
for typical experimental geometries equivalent to illumination and
detection with a plane wave, both for backscattering and
transmission through an optically-thick slab.  However, we find that
the Gaussian character breaks down for point-in / point-out
backscattering geometries in which the illumination spot size is not
sufficiently large in comparison with the size of the intermittent
rearrangement events.
\end{abstract}


submitted to {\it Applied Optics} as a Photon Correlation and
Scattering feature paper

\maketitle 

\section{Introduction}

Aqueous foams consist of a dispersion of gas within a surfactant
stabilized liquid network \cite{AndySciAm89, JoelRPP93, ECT4foam,
WeaireBook99}.  Foams have many practical uses including fire
fighting agents, food and beverage products, and oil recovery. Their
structure is meta-stable, constantly evolving by gas diffusion,
bubble rearrangement, drainage, and bubble rupture.  In fact, if one
waits long enough, the end result for a foam is always the same: one
gas bubble with no intervening liquid.  Foam rheology is robust,
displaying both solid and liquid behavior for small and large sheer
stress, respectively.  In addition, foams present an experimentally
accessible example of a jammed system, an important class of systems
for which the physics is still not fully understood.

Visible light entering a foam is easily scattered due to the index
of refraction mismatch at the liquid and air interfaces. Naturally
occurring foams are disordered, producing randomly oriented
scattering sites. As such, photons perform an effective random walk
as they scatter through the foam \cite{MoinJOSA97, MoinAO01,
AlexEPL04, HolgerPRE03, HolgerEL04}.  This multiple scattering of
light gives foams their white appearance and renders more
traditional observation techniques, such as video imaging,
ineffectual for examining foam structure in the bulk.

Diffusing-wave spectroscopy (DWS) is a general method for taking
advantage of multiple light scattering for a non-invasive probe of
the dynamics of the scattering sites \cite{DWSrev93, DWSrev97,
PierrePRE98}.  As in the more usual single-scattering
photon-correlation technique \cite{BernePecora, ChuBook, BrownBook},
the time trace of the intensity $I(t)=E(t)E^{*}(t)$ of roughly one
speckle of scattered light is measured, and the intensity
autocorrelation is computed:
\begin{eqnarray}
\label{eq:g2}
g^{(2)}(\tau)=\langle{I(0)I(\tau)}\rangle/\langle{I}\rangle^2.
\end{eqnarray}
If the statistics of the underlying electric field are Gaussian,
then the right hand side of this equation is a four-time electric
field correlation that can be expressed as the sum of products of
two-time field correlations.  The result simplifies to the
well-known Siegert relation,
\begin{eqnarray}
\label{eq:Siegert}
  g^{(2)}(\tau) & = & 1 + \beta
  \frac{|\langle E(0) E^{*}(\tau) \rangle|^2}
  {|\langle{EE^{*}\rangle}|^2} \nonumber \\
  & = & 1 + \beta |\gamma(\tau)|^2,
\end{eqnarray}
where $\beta$ is a constant that depends on the ratio of speckle to
detector size \cite{BernePecora, ChuBook, BrownBook}. The Siegert
relation is crucial since it relates the experimentally-measured
intensity fluctuations to the normalized electric field
autocorrelation function, $\gamma(\tau)$, from which the scattering
site dynamics can be determined.

The above procedure has been used in DWS studies of bubble
rearrangements in aqueous foams under several different
circumstances.  For coarsening foams, the diffusion of gas from
smaller to larger bubbles causes sudden stick/slip-like
rearrangement of several neighboring bubbles at a time
\cite{DJDsci91, DJDpra91, AnthonyJOSA97}.  For foams subjected to
steady shear, the macroscopic deformation is accomplished by
microscopic bubble-scale rearrangements \cite{earnshawPRE94,
AnthonyPRL95}.  After cessation of shear, coarsening-induced
rearrangements are suppressed until the bubble-size distribution
noticeably changes by further coarsening \cite{AnthonyPRL95,
hohlerPRL01}.  For oscillating shear, the size of echoes in the DWS
signal can be used as a measure of microscopic reversibility in the
bubble motion \cite{PascalPRL97, hohlerPRL97}.

The Siegert relation requires an assumption that the electric field
$E(t)$ be a Gaussian variable of zero mean. Though this condition is
satisfied for many experimental situations, it can break down if the
scattering sites are few in number, correlated, or if the dynamics
systematically change with time. In addition, some experimental
pitfalls, such as limited laser coherence, or a static component in
the scattered field, can lead to failure.  None of these
possibilities can be eliminated by inspection of $g^{(2)}(\tau)$
alone.

Higher-order statistics can be measured to verify whether or not the
Siegert relation can be invoked to extract the electric field
autocorrelation from the measured intensity autocorrelation
\cite{PierreJOSA99, PierrePRL00, PierreAO01}.  This only involves
further processing of the same bitstream of photon counts.  No
additional auxiliary optical setup is needed, as in heterodyning. In
this paper, we measure the three-time intensity correlation,
simultaneously with $g^{(2)}(\tau)$, for coarsening foams to
determine if the electric field fluctuations are indeed Gaussian, as
assumed in prior studies.  We begin by describing our experimental
methods and applying them to a controlled sample of diffusing
polystyrene spheres, known in advance to possess Gaussian field
fluctuations. After thus verifying our apparatus and procedures, we
investigate the nature of the electric field fluctuations for light
scattered from a coarsening foam.

\section{Optical Methods}

A typical experimental setup for DWS consists of a coherent laser,
the sample, a photon counting device, and a digital correlator. We
use a Coherent Verdi-V5 ($5W$) diode-pumped solid-state laser
operating at $\lambda = 532nm$.  Scattered light is collected into a
photodetector using a single mode optical fiber with GRIN lens and
line filter. The analog photocurrent signal is then amplified and
discriminated such that each photon produces a TTL logic pulse. For
a typical DWS measurement, the resulting bitstream $n(t)$ is sent to
a commercial correlator (Flexible Instruments FLEX1000) for our
experiment, which computes the autocorrelation of the bitstream:
\begin{equation}
\label{eq:g2count}
 g^{(2)}(\tau) =
 \langle{n(0)n(\tau)}\rangle/\langle{n}\rangle^2.
\end{equation}
The right hand side of this equation represents the photon-count
autocorrelation function; $n(t)$ is the number of photons detected
between times $t$ and $t+\tau_s$, where $\tau_s$ is the sampling
time. This is equivalent to the intensity autocorrelation function.

Next, we set out to measure time slices of fixed $T$ of the third
order intensity correlation function:
\begin{equation}
\label{eq:g3}
 g^{(3)}(\tau, \tau+T)  =
 \frac{\langle{I(0)}I(\tau)I(\tau+T)\rangle}{\langle{I}\rangle^3}.
\end{equation}
We supplement the usual DWS set-up, described above, with a
custom-built synchronizing multiplying digital delay line SMDDL. The
same hardware and procedure from our earlier work was
employed\cite{PierreJOSA99, PierrePRL00, PierreAO01}. The delay line
takes the un-synchronized TTL bitstream, $n_u(t)$, and outputs two
synchronized bitstreams: $n_A(t) = n(t)$; and $n_B(t) = n(t)n(t+T)$.
The delay time, $T$, can be set from $50$~ns to about $50$~ms in
increments of $50$~ns.  The two synchronized bitstreams $n_A(t)$ and
$n_B(t)$ are fed to the digital correlator as inputs.
Auto-correlating channel $A$ produces $g^{(2)}(\tau)$ given by Eq.
(\ref{eq:g2count}). The cross-correlation of channel $A$ and $B$,
$\langle{n_A(0)n_B(\tau)}\rangle/(\langle{n_A}\rangle\langle{n_B}\rangle)$,
produces a slice of the three-time correlation:
\begin{equation}
\hat{g}^{(3)}(\tau, \tau+T)  =
\frac{\langle{n(0)}n(\tau)n(\tau+T)\rangle}
{\langle{n_A}\rangle\langle{n_B}\rangle},
\end{equation}
where $T$ and $\tau$ are the delays introduced by the SMDDL and the
correlator, respectively, and $\langle{n_A}\rangle$ and
$\langle{n_B}\rangle$ are the average number of counts per sampling
time for channels $A$ and $B$. The raw correlation
$\hat{g}^{(3)}(\tau, \tau+T)$ does not have the correct
normalization for comparison with ${g}^{(3)}(\tau, \tau+T)$. This
can be recovered by noting that $\langle{n_A}\rangle =
\langle{n}\rangle = R_A\tau_s$ and $\langle{n_B}\rangle =
\langle{n(\tau)n(\tau+T)}\rangle = R_B\tau_s$, where $R_A$ and $R_B$
are the count rates measured by correlator channels $A$ and $B$,
respectively.  Properly normalized, we have
\begin{eqnarray}
 {g}^{(3)}(\tau, \tau+T) & = &
\label{eq:g3count}
 \frac{\langle{n_B}\rangle}{\langle{n_A}\rangle^2}\hat{g}^{(3)}(\tau, \tau+T) =
 \frac{\langle{n(0)}n(\tau)n(\tau+T)\rangle}{\langle{n}\rangle^3},  \\
 & = &
\label{eq:g3data}
 \frac{R_{B}}{\tau_s R_A^2}\hat{g}^{(3)}(\tau, \tau+T).
\end{eqnarray}
The three-time counts correlation function, given by the right hand
side of Eq.(\ref{eq:g3count}), is equivalent to the three-time
intensity correlation function Eq.(\ref{eq:g3}) since each photon
count produces a digital TTL pulse. The right hand side of
Eq.(\ref{eq:g3data}) represents the experimental measurement of
${g}^{(3)}(\tau, \tau+T)$.  Altogether, we are thus able to measure
both $g^{(2)}(\tau)$ and a constant-$T$ slice of ${g}^{(3)}(\tau,
\tau+T)$ simultaneously.

Next, we generate the three-time Gaussian prediction
${g}^{(3)}_{G}(\tau, \tau+T)$.  The first step is to extract
$\gamma(\tau)$ and $\beta =
\langle{I^2}\rangle/\langle{I}\rangle^2-1$ from $g^{(2)}(\tau)$,
assuming that the Siegert relation Eq.(\ref{eq:Siegert}) is valid.
For Gaussian scattering processes, intensity correlation functions
of any order can be expressed in terms of sums of products of the
electric field autocorrelation function. The three-time intensity
correlation function is a six-time field correlation function that,
if Gaussian, reduces to\cite{PierreJOSA99}
\begin{eqnarray}
\label{eq:g3Gauss}
 {g}^{(3)}_{G}(\tau, \tau+T) = 1 +
 \beta[|\gamma(\tau)|^2 + |\gamma(T)|^2 + |\gamma(\tau+T)|^2] +
 2\beta^{2}
 \begin{mathrm} Re
 \end{mathrm}[\gamma(\tau)\gamma(T)\gamma(\tau+T)].
\end{eqnarray}
Similar predictions have appeared in prior literature
\cite{DegiorgioPRA76,PhilliesJCP80,SchurrCP91}; however none
considered the nonzero detector area as accounted for here by
factors of $\beta$.

For our measurements we employ two illumination and collection
geometries: plane-in / plane-out equivalent \cite{DJDA095}, and
point-in / point-out. For both geometries, the sample is contained
within a glass cell, and the incident beam is normal to the sample
surface. Plane-in illumination is accomplished by expanding the
laser beam with a diverging lens to be greater than the cell
thickness. Point-in illumination is accomplished by passing the beam
through a converging lens one focal length from the sample; the
resulting spot size is about 0.5~mm. In all cases (including
plane-equivalent \cite{DJDA095}), the detection spot size is
slightly larger than the diameter of the GRIN lens, about 2~mm.

\section{Diffusing particles}

First we test our apparatus and procedures on diffusing Brownian
particles.  Our sample consists of polystyrene spheres, diameter
$d=93\pm8$~nm, suspended in a 53\% glycerol solution at a volume
fraction of 1.86\%, and contained in a 9~mm thick glass cell.  This
sample was previously measured at a slightly different wavelength
\cite{PierrePRE98}. At the wavelength used here, the average cosine
of the scattering angle is $g=0.102$ and the scattering mean free
path is $l_{s}=0.533$~mm; these give the transport mean free path
$l^{*}=l_{s}/(1-g)=0.594$~mm.  There is negligible absorption.  The
characteristic diffusion time that enters DWS predictions is
$\tau_\circ = 1/(Dk^2)=5.90$~ms, where $D$ is the particle diffusion
coefficient and $k$ is wavenumber of the laser light in the
solution.

\begin{figure}[htbp]
\centering
\includegraphics[width=8.3cm]{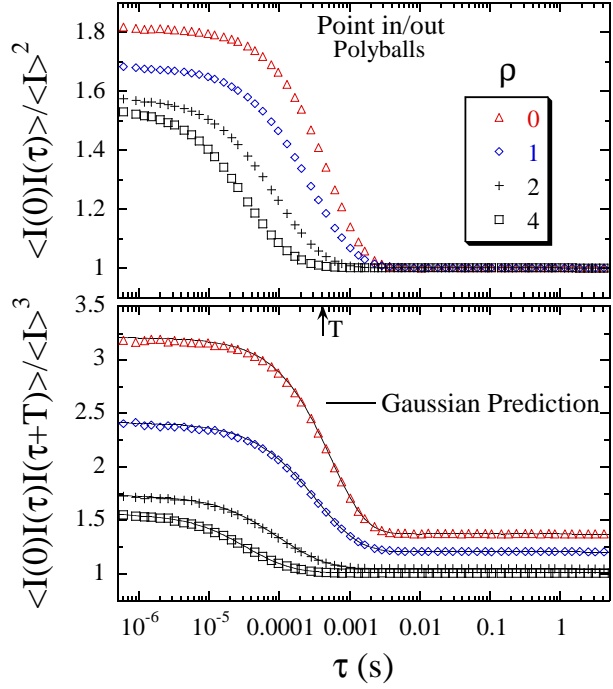}
\caption{(color online) Two- and three-time intensity correlation
functions for light backscattered from a colloidal suspension of
polystyrene particles with a point-in / point-out geometry. Data are
shown by symbols, while the Gaussian predictions based on
autocorrelation data are shown by solid curves. The separation
distance between the centers of the illumination and detection spots
is given by $\rho$ in units of the transport mean free path, $l^*$,
as labeled. The fixed delay is $T=0.410$~ms, as marked by an arrow.}
\label{PointPoly}
\end{figure}

Intensity autocorrelation results for a point-in / point-out
backscattering geometry are shown in the top plot of
Fig.~\ref{PointPoly}. The separation distances between entry and
collection spots, measured in units of $l^*$, for four different
runs are $\rho = \{0, 1, 2, 4\}$. If the illumination and detection
spots are small and if their separation is large, all in comparison
with $l^{*}$, then the DWS prediction is $\gamma(\tau) =
(1+\rho\sqrt{6\tau/\tau_\circ}) \exp(-\rho\sqrt{6\tau/\tau_\circ})$
\cite{MorinAO02}.  While these conditions do not hold for our
experiments, we nevertheless find excellent fits to this form by
adjusting the value of $\rho$.  The resulting effective
illumination-detection separation distances, in units of $l^*$, for
our four runs are $\{1.8, 2.4, 4.1, 6.6\}$.  To accurately model the
intensity autocorrelation data, using the theory of DWS, would
require knowledge of intensity and sensitivity vs position for the
illumination and detection spots, respectively; however, this is not
necessary for our main purpose of testing the validity of the
Siegert relation.

Three-time intensity correlation results for point-in / point-out
backscattering are shown in the bottom plot of Fig.~\ref{PointPoly}.
These data were collected simultaneously with the intensity
autocorrelation data in the top plot, where the same constant delay
time $T=0.410$~ms was used for all four runs.  To test whether the
field statistics are Gaussian, we first extract the field
autocorrelation assuming Eq.~(\ref{eq:Siegert}). Then we use it to
generate the Gaussian expectation for $g^{(3)}(\tau,\tau+T)$ using
Eq.~(\ref{eq:g3Gauss}). Finally we plot this expectation along with
the actual data in Fig.~\ref{PointPoly}b.  Evidently, the Gaussian
expectation agrees very well with the data.  Therefore, we conclude,
the electric field truly does have Gaussian statistics.  This was
expected, by design, because there are large numbers of uncorrelated
diffusing particles within the scattering volume.  This
demonstration validates our experimental apparatus and methods.  We
now proceed to apply the same methods to a coarsening foam, which is
not known in advance to be Gaussian.

\section{Coarsening Foam}

Our sample consists of Gillette Foamy shaving cream, composed by
volume of 92\% polydisperse gas bubbles tightly packed in an aqueous
surfactant solution.  With time, this foam coarsens due to the
diffusion of gas from smaller to larger bubbles; drainage and film
rupture are negligible. The sample container is a $L=7$~mm thick
glass cell with much larger lateral dimensions, so that light does
not escape from the sides. All light scattering measurements are
performed after the foam has aged for 100~minutes in the sample
cell. At this age, the average bubble diameter is $d=60$~$\mu$m and
the photon transport mean free path is $l^*=210$~$\mu$m
\cite{DJDsci91}. The measurement duration is 500~s, which is long
enough to get a good measure of the intercept and baseline of the
intensity autocorrelation, but small enough that the bubble size
distribution does not significantly change.

The speckle pattern formed by scattered light fluctuates with time
due to sudden structural rearrangements of small groups of
neighboring bubbles from one packing configuration to another. These
events relax local stress inhomogeneities that accumulate due to the
coarsening process.  The duration of a typical event is a few tenths
of a second, while the time between successive events at any given
location is roughly $\tau_\circ=20$~s, for our 100~minute old
samples. Thus, the bubbles are usually static and they rearrange
only intermittently and quickly. Such dynamics are very different
from thermal Brownian motion, where all particles gradually and
independently move around.  Therefore, for foam, the electric field
statistics could conceivably be non-Gaussian either because most of
the scattering sites are static, and hence correlated, or because
the motion during an event is correlated throughout a scattering
volume. Nevertheless, prior analysis has always assumed Gaussian
field fluctuations that appear exactly like Brownian diffusion.  The
only difference is that now $\tau_\circ=20$~s represents the time
between rearrangement events at a given scattering site, as opposed
to the time required for a particle to diffuse one wavelength.

\begin{figure}[htbp]
\centering
\includegraphics[width=8.3cm]{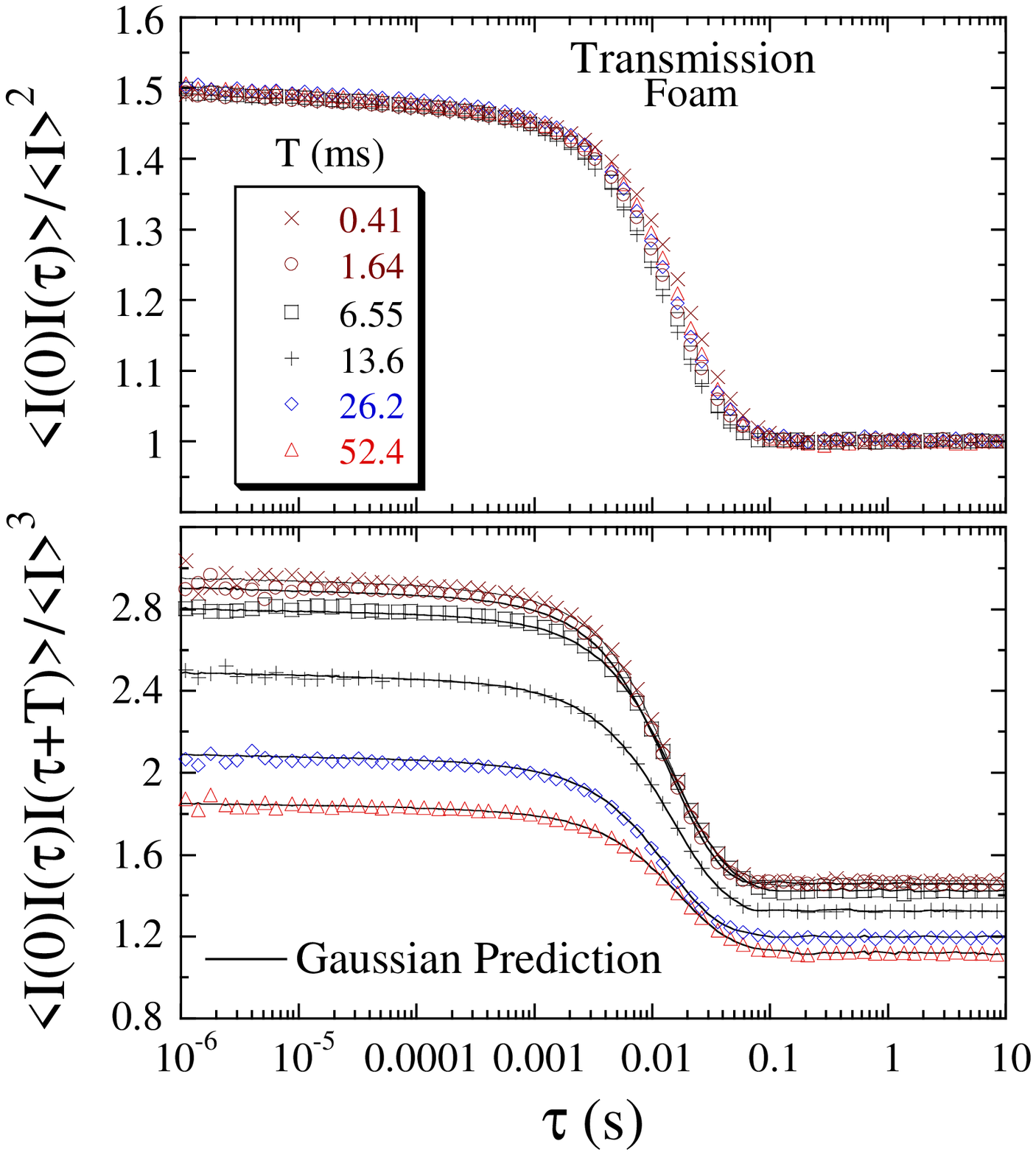}
\caption{(color online) Two- and three-time intensity correlation
functions for light transmitted through foam with a plane-in /
plane-out equivalent geometry. Data are shown by symbols, while the
Gaussian predictions based on autocorrelation data are shown by
solid curves. The fixed delay time $T$ is different for each of the
six runs, as labeled.} \label{TransFoam}
\end{figure}

Intensity autocorrelation results for a plane-in / plane-out
transmission geometry are shown in the top plot of
Fig.~\ref{TransFoam}. At short times there is a slight gradual decay
due to thermal motion of the bubble interfaces \cite{AnthonyJOSA97}.
At long times there is a full decay due to rearrangements that is
well-described by $\gamma(\tau) = \sqrt{(L/l^*)^2 6\tau/\tau_\circ}
/ \sinh[\sqrt{(L/l^*)^2 6\tau/\tau_\circ}]$ \cite{DJDsci91}.
Three-time intensity correlation data were acquired simultaneously
for six different fixed delays, $T$, as labeled; the results are
shown in the bottom plot of Fig.~\ref{TransFoam}.  The Gaussian
expectation is generated directly from the intensity autocorrelation
data, using Eqs.~(\ref{eq:Siegert}, \ref{eq:g3Gauss}), and the
results are plotted along with the actual data
Fig.~\ref{TransFoam}b. Evidently the agreement is very good,
justifying the long-standing assumption of Gaussian fluctuations for
this geometry.

\begin{figure}[htbp]
\centering
\includegraphics[width=8.3cm]{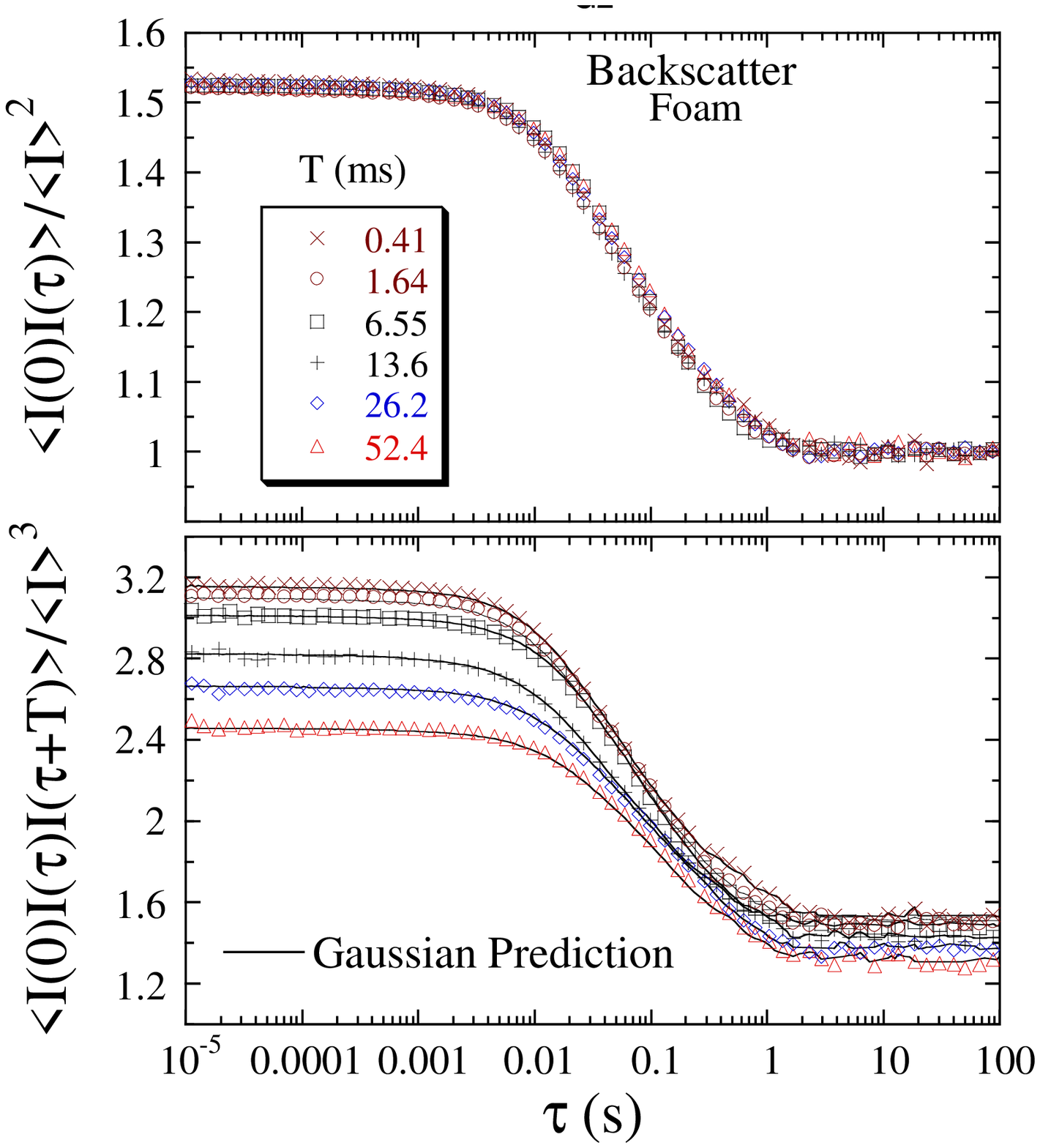}
\caption{(color online) Two- and three-time intensity correlation
functions for light backscattered from foam with a plane-in /
plane-out equivalent geometry. Data are shown by symbols, while the
Gaussian predictions based on autocorrelation data are shown by
solid curves.  The fixed delay time $T$ is different for each of the
six runs, as labeled.} \label{BackFoam}
\end{figure}

Next we carry out the same program for a plane-in / plane-out
backscattering geometry.  Since backscattering features shorter
light paths than in transmission, there could now be non-Gaussian
fluctuations. The results for the two- and three-time intensity
correlations at six different fixed delay times, and the Gaussian
expectation generated from the autocorrelation data, are all shown
in Fig.~\ref{BackFoam}. The field correlation function is
well-described by $\gamma(\tau) = \exp(-2\sqrt{6\tau/\tau_\circ})$
\cite{DJDsci91}.  The agreement between the Gaussian expectation and
the actual three-time data is very good, again justifying the
long-standing assumption of Gaussian fluctuations for this geometry.

\begin{figure}[htbp]
\centering
\includegraphics[width=8.3cm]{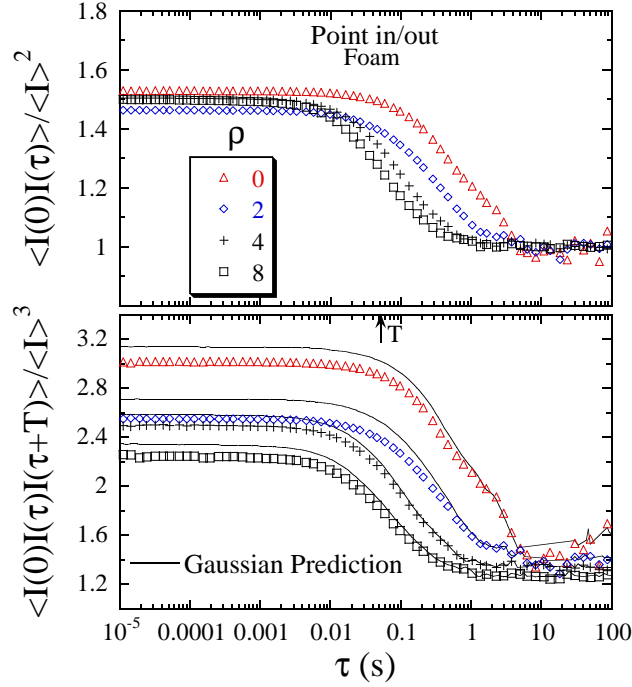}
\caption{(color online) Two- and three-time intensity correlation
functions for light backscattered from a foam with a point-in /
point-out geometry. Data are shown by symbols, while the Gaussian
predictions based on autocorrelation data are shown by solid curves.
The separation distance between the centers of the illumination and
detection spots is given by $\rho$ in units of the transport mean
free path, $l^*$, as labeled. The fixed delay is $T=52.4$~ms, as
marked by an arrow.} \label{PointFoam}
\end{figure}

Finally we repeat the same program for point-in / point-out
backscattering geometries with four different separation distances.
The results for the two- and three-time intensity correlations at
the same fixed delay time, and the Gaussian expectation generated
from the autocorrelation data, are all shown in
Fig.~\ref{PointFoam}.  By contrast with our other measurements, now
we find non-Gaussian field fluctuations: the Gaussian expectation is
systematically higher than the actual three-time intensity
correlation data. Hence, the intensity autocorrelation is not
described by the usual Siegert relation, Eq.~(\ref{eq:Siegert}), and
the field correlation is not described by $\gamma(\tau) =
(1+\rho\sqrt{6\tau/\tau_\circ}) \exp(-\rho\sqrt{6\tau/\tau_\circ})$,
as they were for the polyball sample of Fig.~\ref{PointPoly}.

Because of the non-Gaussian character, there must be information
available in the three-time intensity correlation that is not
present in the intensity autocorrelation alone.  But what
information would this be?  One possibility is fluctuations in the
number of scattering sites within the scattering volume.  There have
been many demonstrations of how number fluctuations give rise to
non-Gaussian effects \cite{Estes71,Pusey79,PierreJOSA99}.  However,
here we find Gaussian statistics for a polyball sample that has a
larger $l^*$ by a factor of two.  Thus we believe that the extra
information for a coarsening foam has to do with the intermittent
nature of the rearrangement dynamics.  One possibility is that the
non-Gaussian character could be analyzed in terms of the
rearrangement event size in comparison with the illumination /
detection spot sizes.  Another possibility is that it could be
analyzed in terms of the switching functions that describe the
statistics for how rearrangement events start and stop, as was done
earlier for avalanches in intermittently-flowing sand
\cite{PierrePRL00,PierreAO01}.  In either case, by contrast with
number fluctuations, the non-Gaussian character is due to scattering
site dynamics and is not evident in the intensity distribution.
These possibilities could be investigated by multispeckle techniques
such as TRC\cite{TRC,TRConFoam} or SVS\cite{SVSPRL,SVSRSI}.

\section{Conclusion}

Higher-order temporal intensity correlation measurements are a
powerful tool to confirm or deny the Gaussian character of electric
field statistics.  As applied to a coarsening foam, we find that
prior assumptions of Gaussian fluctuations were indeed warranted;
this was not obvious in advance. The only proviso is that the spot
sizes in plane-in / plane-out equivalent geometries~\cite{DJDA095}
be made sufficiently large in comparison with the rearrangement
event size.  The origin of non-Gaussian fluctuations for small spots
sizes is likely due to the intermittent nature of the scattering
site dynamics, by contrast with the well-known case of number
fluctuations \cite{Estes71,Pusey79,PierreJOSA99}, and warrants
further study.

\section*{Acknowledgments}
We thank Pierre-Anthony Lemieux for technical assistance. This work
was supported by NASA under Microgravity Fluid Physics Grant
NAG3-2481.





\newpage

\section*{List of Figure Captions}

\noindent Fig. 1. (color online) Two- and three-time intensity
correlation functions for light backscattered from a colloidal
suspension of polystyrene particles with a point-in / point-out
geometry. Data are shown by symbols, while the Gaussian predictions
based on autocorrelation data are shown by solid curves. The
separation distance between the centers of the illumination and
detection spots is given by $\rho$ in units of the transport mean
free path, $l^*$, as labeled. The fixed delay is $T=0.410$~ms, as
marked by an arrow.

\vfil

\noindent Fig. 2. (color online) Two- and three-time intensity
correlation functions for light transmitted through foam with a
plane-in / plane-out equivalent geometry. Data are shown by symbols,
while the Gaussian predictions based on autocorrelation data are
shown by solid curves. The fixed delay time $T$ is different for
each of the six runs, as labeled.

\vfil

\noindent Fig. 3. (color online) Two- and three-time intensity
correlation functions for light backscattered from foam with a
plane-in / plane-out equivalent geometry. Data are shown by symbols,
while the Gaussian predictions based on autocorrelation data are
shown by solid curves. The fixed delay time $T$ is different for
each of the six runs, as labeled.

\vfil

\noindent Fig. 4. (color online) Two- and three-time intensity
correlation functions for light backscattered from a foam with a
point-in / point-out geometry. Data are shown by symbols, while the
Gaussian predictions based on autocorrelation data are shown by
solid curves. The separation distance between the centers of the
illumination and detection spots is given by $\rho$ in units of the
transport mean free path, $l^*$, as labeled. The fixed delay is
$T=52.4$~ms, as marked by an arrow.

\vfil

\newpage

\end{document}